\documentclass[12pt,a4wide]{article}

\usepackage[latin1]{inputenc}
\usepackage{amsmath}
\usepackage{amsfonts}
\usepackage{amssymb}
\usepackage{graphicx}
\usepackage{epstopdf}
\usepackage{color}
\usepackage{appendix}
\usepackage{csvsimple}
\usepackage{longtable}
\usepackage{chngpage}
\usepackage{subcaption}
\usepackage[export]{adjustbox}[2011/08/13]
\usepackage{mathrsfs}
\usepackage{a4wide}
\usepackage{psfrag}
\usepackage{multirow}
\usepackage{hhline}
\usepackage{grffile}

\newcommand{\lsim}{\mbox{\raisebox{-.9ex}{~$\stackrel{\mbox{$<$}}{\sim}$~}}}

\begin{document}

\begin{center}
  {\large\bf
Waterfall stiff period can generate observable primordial gravitational waves}

\bigskip

{\large Konstantinos Dimopoulos}\footnote{\tt k.dimopoulos1@lancaster.ac.uk}

\bigskip

{\em Consortium for Fundamental Physics,
  Physics Department,\\
  Lancaster University, Lancaster LA1 4YB, UK}
\end{center}

\begin{abstract}
  A toy-model is studied, which considers two flat directions meeting at an
  enhanced symmetry point such that they realise the usual hybrid inflation
  mechanism. The kinetic term of the waterfall field features a pole at its
  Planckian vacuum expectation value (VEV), as with $\alpha$-attractors.
  Consequently, after the phase transition which terminates hybrid inflation,
  the waterfall field never rolls to its VEV. Instead, it drives a stiff period,
  where the barotropic parameter of the Universe $w\approx 1/2$ results in a
  peak in the spectrum of primordial gravitational waves, which will be
  observable by the forthcoming LISA mission as well as by Advanced LIGO.
\end{abstract}

\section{Introduction}
The most compelling origin story for our Universe is cosmic inflation,
which, not only solves in a single stroke the fine-tuning problems of the
Hot Big Bang cosmology (the horizon and flatness problems) but neatly generates
the primordial density perturbations necessary for the eventual formation
of structures in the Universe, such as galaxies and galactic clusters
\cite{starobinsky,guth} \cite{kazanas,sato}. In fact,
after the observations of the CMB acoustic peaks which lead to the collapse of
the rival paradigm of cosmic strings \cite{rise&fall} for structure formation,
cosmic inflation is considered as an essential extension of the Hot Big Bang,
in the cosmological standard model \cite{lythbook}.

The acoustic peaks, even though a prediction of inflation, were not thought to
be a smoking gun. This is reserved for another generic prediction of inflation,
that of primordial gravitational waves. Indeed, in a similar manner to the way
inflation generates the density perturbations, it
is also expected to result in a flat spectrum of primordial gravitational waves
\cite{turnerGW}. As we have entered a new era of
gravitational wave astronomy, observing these gravitational waves is of
paramount importance, which is expected to cement inflation as the necessary
extension of the Hot Big Bang.

Unfortunately, the amplitude of the inflation generated primordial gravitational
waves is typically too small to observe in the near future, by Advanced LIGO
\cite{LIGO}, Virgo \cite{Virgo} or the space interferometer LISA \cite{LISA}
(see also Ref.~\cite{LIGO3Virgo}). Yet, there are certain types of inflation,
namely non-oscillatory (NO) inflation \cite{NOinflation}, which may offer this
possibility.\footnote{NO inflation is called so because it does not involve
  oscillations of the inflaton field after inflation ends. Reheating in NO
  models is achieved if a way other than the inflaton field decay. NO models are
  frequently employed in quintessential
  inflation \cite{QI}. For recent reviews see Refs.~\cite{samiQI,haroQI}.}
This is because, in these models, the spectrum on primordial
gravitational waves, apart from the almost scale-invariant plateau, can feature
a peak of enhanced gravitational waves \cite{sidorov,sahni},
large enough to render them observable \cite{steinhardtGW,saito,kamionkowski}.
The reason for this is the following.

The (almost) flat primordial gravitational wave spectrum corresponds to the
scales which, after exiting the horizon during quasi-de Sitter inflation,
re-enter the horizon during the radiation dominated period of the Hot Big Bang.
Because the densities of the thermal bath of the Hot Big Bang and of
gravitational radiation are decreasing in time equally fast, there is no
difference when a particular scale (mode) re-enters the horizon. This is why
the spectrum is predominantly flat. However, in NO inflation models, there is
a possibility that, before reheating and the radiation era, the Universe is
dominated by the kinetic energy density of the inflaton field, resulting in
a period called kination \cite{kination}.\footnote{In general, `kination'
  refers to any period when the Universe is dominated by a substance with
  barotropic parameter \mbox{$w=1$}, for example a minimally coupled scalar
  field dominated by its kinetic energy density.}
The equation of state of the Universe during kination
is stiff, with a barotropic parameter $w=1$. The density of stiff matter
redshifts faster than the density of gravitational radiation, so the spectrum of
primordial gravitational waves is no-longer flat, for the modes corresponding
to the scales which re-enter the horizon during kination, but it gives rise to
a peak of enhanced gravitational radiation.\footnote{Incidentally,
  gravitational waves from
  kination can also alleviate the Hubble tension \cite{GW&HubbleTension}.}

However, this possibility suffers also from a big problem. Kination typically
follows the end of inflation in NO models. This means that the peak in the
spectum of the primordial gravitational waves corresponds to very high
frequencies, because the inflation energy scale is typically very high (near the
energy of grand unification). The more kination lasts, the lower the frequencies
that the peak in the gravitational wave spectrum
extends to. Unfortunately, kination cannot be made to last enough so that the
enhancement includes observable scales. The reason is that such a long kination
period would result is an exceptionally large peak corresponding to a huge
energy density of primordial gravitational radiation, which would be so large
as to destabilise the sacred cow of Hot Big Bang cosmology, the process of
Big Bang Nucleosynthesis (BBN). Thus, making sure that BBN is not disturbed,
means that kination cannot last too long and the peak in the spectrum of
primordial
gravitational waves is confined to frequencies too large to be observable in the
near future \cite{haroBBN}.

Yet, there is a way out. If the barotropic parameter of the stiff era is not
$w=1$ but assumes a value in the range \mbox{$1/3<w<1$} then there will still be
growth of gravitational radiation but it will not be so sharply peaked as in the
case of kination, with $w=1$. As a result, the stiff period could be extended
to lower frequencies without the peak in the spectrum of primordial
gravitational waves becoming forbiddingly large.\footnote{This is the simplest
  but by no means the only possibility. For example, the stiff period may not
  begin right after inflation but later on.} In the recent
work in Ref.~\cite{FigTanin}, it was shown that, if the barotropic parameter of
the stiff era lies in the range \mbox{$0.46\lsim w\lsim 0.56$} and the reheating
temperature at the beginning of radiation domination is
\mbox{1\ MeV$\,\lsim T_{\rm reh}\lsim 150\,$MeV}, primordial gravitational waves 
can be enhanced enough to become observable by LISA without disturbing BBN.
But how can such a stiff era be generated?

In this paper we provide a toy-model realisation of this possibility. We
consider two flat directions is field space which cross each other at an
enhanced symmetry point (ESP). One of these flat directions can play the role
of the inflaton field, while the other one, which develops a tachyonic mass at
the ESP, can be the waterfall field in a classic hybrid inflation setup
\cite{hybrid}. The waterfall field vacuum expectation value (VEV) is Planckian,
so that, after the rolling inflaton reaches the ESP, a phase transition
terminates primordial inflation and sends the system rolling along the waterfall
direction, which however results in a small number of e-folds of hilltop
fast-roll inflation. The crucial element in our model is that the kinetic term
of the waterfall field is non-canonical, but instead features a pole at the VEV
of the waterfall field, as with $\alpha$-attractors \cite{alpha}.
Consequently, as the system moves away from the ESP, the
dynamics of the rolling waterfall field are modified and the VEV is never
reached. Therefore, this is a NO inflation scenario.

After the end of the hilltop fast-roll inflation period, the waterfall field
continues to roll, but not quite dominated by its kinetic energy density.
Instead, it is following an attractor solution which corresponds to a stiff
barotropic parameter but smaller that unity. We show that the value of the
barotropic parameter is determined by the waterfall field VEV. Hence, this VEV
can be tuned to fall into the region \mbox{$0.46\lsim w\lsim 0.56$} such that
we can obtain observable gravitational waves generated by primordial inflation
\cite{FigTanin}.

In the following, we use natural units where \mbox{$c=\hbar=k_B=1$} and
\mbox{$8\pi G=m_P^{-2}$}, with \mbox{$m_P=2.43\times 10^{18}\,$GeV} being the
reduced Planck mass. 

\section{The model}

Consider a theory with Lagrangian density
\mbox{${\cal L}={\cal L}_{\rm kin}-V$},
where the scalar potential is
\begin{equation}
  V(\varphi,\sigma)=\frac12 g^2\sigma^2\varphi^2+
  \frac14\lambda\left(\varphi^2-M^2\right)^2+V(\sigma)
\label{Vvarphis}
\end{equation}  
and the kinetic Lagrangian density is
\begin{equation}
  {\cal L}_{\rm kin}=\frac12(\partial\sigma)^2+
  \frac{\frac12(\partial\varphi)^2}{\left(1-\varphi^2/M^2\right)^2}\,,
\label{Lkin}
\end{equation}
where \mbox{$(\partial\sigma)^2=-\partial_\mu\sigma\,\partial^\mu\sigma$}
and \mbox{$(\partial\varphi)^2=-\partial_\mu\varphi\,\partial^\mu\varphi$}
with metric signature ($-$,+,+,+). 

\begin{figure} [t]
\centering
\includegraphics[width=10cm]{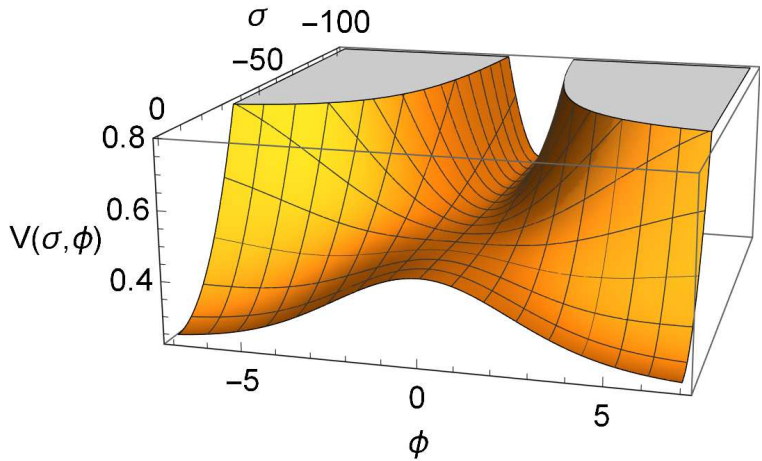} 
\caption{Qualitative form of the scalar potential in Eq.~(\ref{Vphis}) in terms
  of the canonical scalar fields $\sigma$ (inflaton) and $\phi$ (waterfall). The
  axes are in fiducial units. The system origitally finds itself gradually
  rolling $V(\sigma)$ inside the steep valey at $\phi=0$ when $\sigma$ is large.
  When the inflaton is reduced below the value $\sigma_c$, given in
  Eq.~(\ref{sigmac}), a phase transition sends the waterfall field away from the
  origin. The minima along the canonical waterfall direction are displaced at
  infinity. As a result, after the phase transition, there is an initiall period
  of fast-roll hilltop inflation along the waterfall direction (discussed in
  Sec.~\ref{watinf}), followed by a stiff period (discussed in
  Sec.~\ref{watstiff}).}  
\label{3dplot}
\end{figure}

In the above, the scalar potential is in the standard form of the hybrid
mechanism \cite{hybrid}. The scalar field $\sigma$ is the inflaton field, while
$\varphi$ is the waterfall field. $V(\sigma)$ is the inflaton potential.
However, the kinetic term of the $\varphi$
scalar field features poles at \mbox{$\varphi=\pm M$}, which can be motivated
in conformal field theory or in supergravity with a non-trivial K\"{a}hler
manifold. This is the basis of $\alpha$-attractors \cite{alpha}.\footnote{For
  a recent implementation of $\alpha$-attractors to hybrid inflation see
  Ref.~\cite{LindeRecent}.}
The above suggests that the mass scale $M$ is linked with the $\alpha$
parameter of $\alpha$-attractors as
\begin{equation}
M=\sqrt{6\alpha}\,m_P\;.
\label{Malpha}
\end{equation}
To assist our intuition, we switch to a canonically normalised scalar field
$\phi$, which is related with the non-canonical $\varphi$ as
\begin{equation}
  \frac{{\rm d}\varphi}{1-\varphi^2/M^2}={\rm d}\phi\;\Rightarrow\;
\varphi=M\tanh(\phi/M)\,.
\label{phivarphi}
\end{equation}
Then, the scalar potential, in terms of canonical fields, becomes
\begin{equation}
V(\phi,\sigma)=\frac12\,g^2M^2\sigma^2\tanh^2(\phi/M)+
\frac{\frac14\lambda M^4}{\cosh^4(\phi/M)}+V(\sigma)\,.
\label{Vphis}
\end{equation}
The above potential is shown in Fig.~\ref{3dplot}.

\section{The modified hybrid mechanism}

The first task is to investigate whether the hybrid mechanism operates as usual,
under the new form of the scalar potential. It is straightforward to find
\begin{equation}
\frac{\partial V}{\partial\phi}=M\frac{\sinh(\phi/M)}{\cosh^3(\phi/M)}
\left[g^2\sigma^2-\frac{\lambda M^2}{\cosh^2(\phi/M)}\right]
\label{DV}
\end{equation}
and
\begin{equation}
\frac{\partial^2 V}{\partial\phi^2}=g^2\sigma^2
\frac{1-2\sinh^2(\phi/M)}{\cosh^4(\phi/M)}-\lambda M^2
\frac{1-4\sinh^2(\phi/M)}{\cosh^6(\phi/M)}\,.
\label{DDV}
\end{equation}  

At the origin \mbox{$\phi=0$} we have
\begin{equation}
m_{\rm eff}^2\equiv\left.\frac{\partial^2 V}{\partial\phi^2}\right|_{\phi=0}=
  g^2\sigma^2-\lambda M^2=g^2(\sigma^2-\sigma_c^2)\,,
\label{DDV0}  
\end{equation}
where
\begin{equation}
\sigma_c\equiv\frac{\sqrt\lambda}{g}\,M\,.
  \label{sigmac}
\end{equation}
Thus, we see that \mbox{$m_{\rm eff}^2>0$} (\mbox{$m_{\rm eff}^2<0$}) provided
\mbox{$\sigma>\sigma_c$} (\mbox{$\sigma<\sigma_c$}). Now, Eqs.~(\ref{DV}) and
(\ref{sigmac}) suggest that
\begin{equation}
\frac{\partial V}{\partial\phi}=g^2M\frac{\sinh(\phi/M)}{\cosh^3(\phi/M)}
\left[\sigma^2-\frac{\sigma_c^2}{\cosh^2(\phi/M)}\right]\,.
\end{equation}  
Because \mbox{$\cosh(\phi/M)\geq 1$}, we see that, when
\mbox{$\sigma>\sigma_c$}, the term in the square brackets above is always
positive. This means that the potential in the $\phi$-direction has only one
extremum (where \mbox{$\partial V/\partial\phi=0$}) when \mbox{$\phi=0$}.
Because of Eq.~(\ref{DDV0}), we find that, when \mbox{$\sigma>\sigma_c$}, 
the potential in the $\phi$-direction has a minimum at \mbox{$\phi=0$}, as in
standard hybrid inflation.

Thus, we see that, provided we begin with \mbox{$\sigma>\sigma_c$}, the system
is driven to the valley at \mbox{$\phi=0$}. If the inflaton potential
$V(\sigma)$ provides a gentle slope such that the value of $\sigma$ gradually
diminishes, then at some point the inflaton decreases down to $\sigma_c$, where
the effective mass of the waterfall field becomes tachyonic and we have a phase
transition which terminates inflation in the $\sigma$-direction. The story is
identical with standard hybrid inflation \cite{hybrid}.

\section{Waterfall inflation}\label{watinf}

After the phase transition, the expectation value of the waterfall field
increases. As a result, the interaction term between the two fields becomes a
mass term for the inflaton, which sends it to zero, which presumably also
eliminates the inflaton potential $V(\sigma)$, i.e. assuming
\mbox{$V(\sigma=0)=0$}. Then, Eq.~(\ref{Vphis}) suggests that the potential
becomes
\begin{equation}
V(\phi)=\frac{\frac14\lambda M^4}{\cosh^4(\phi/M)}\,.
\label{Vphi}
\end{equation}
As the waterfall field rolls down the above potential, is gives rise to a bout
of inflation, followed by a stiff period (see next section). Without loss of
generality, we assume that the waterfall field is positive.

Inflation takes place near the hilltop, with \mbox{$0<\phi<M$}.
The potential is approximated as
\begin{equation}
V(\phi)\simeq\frac{\frac14\lambda M^4}{\left[1+\frac12(\phi/M)^2\right]^4}
\simeq\frac14\lambda M^4\left[1-2(\phi/M)^2\right]\,.
\label{VFR}
\end{equation}  
This suggests that this period of hilltop inflation ends when
\mbox{$\phi_{\rm end}\simeq M/\sqrt 2$} \cite{hilltop}.
The penultimate equation in the above,
estimates that the potential density when \mbox{$\phi=\phi_{\rm end}$} has
decreased by a factor \mbox{$~(4/5)^4\approx 0.4$}.

As we discuss in the next session, \mbox{$M\sim m_P$}, which suggests that
the waterfall field undergoes fast-roll inflation \cite{fastroll}.
The reason is that, for the $\eta$ slow-roll parameter, we have
\begin{equation}
  |\eta|=\frac13\frac{|m_{\rm eff}^2|}{H_{P/T}^2}\simeq 
  \frac{\lambda M^2}{\frac14\lambda M^4/m_P^2}=4\left(\frac{m_P}{M}\right)^2\,,
\end{equation}
which is of order unity when \mbox{$M\sim m_P$}. In the above, we used that at
the top of the potential, where the waterfall field finds itself at the phase
transition, we have
\begin{equation}
V(\phi=0)=\frac14\lambda M^4\simeq 3H_{P/T}^2m_P^2\;\Rightarrow\;
H_{P/T}^2\simeq\frac{\lambda M^4}{12\,m_P^2}\,.
\label{HPT}
\end{equation}  

The total e-folds of fast-roll inflation are \cite{fastroll}
\begin{equation}
  N_{\rm FR}=\frac{1}{2F}\ln\left(\frac{\phi_{\rm end}^2}{\phi_{\rm beg}^2}\right)
  \simeq\frac{-\ln(2\lambda)}{2F}\,,
\label{NFR}
\end{equation}  
where we estimated the initial value of the waterfall field as
\mbox{$\phi_{\rm beg}^2\simeq|m_{\rm eff}^2|=\lambda M^2$} and
\begin{equation}
F\equiv\frac32\left(\sqrt{1+\frac43|\eta|}-1\right)\,.
\label{F}  
\end{equation}

$N_{\rm FR}$ can be large if \mbox{$\lambda\ll 1$}. We can estimate $\lambda$ as
follows. The potential density on top of the hill is the same as in the valley
of the hybrid potential, which is the one that drives primordial inflation
along the $\sigma$ direction. Typically, in order to obtain the correct
amplitude for the curvature perturbation, the potential density of primordial
inflation is \mbox{$V_{\rm inf}\sim 10^{-10}\,m_P^4$}. Thus we find
\begin{equation}
\frac14\lambda M^4\simeq V_{\rm inf}\simeq 10^{-10}\,m_P^4\;\Rightarrow\;
  \lambda\simeq 4\times 10^{-10}\left(\frac{m_P}{M}\right)^4\,.
\label{lambda}
\end{equation}

\section{Waterfall stiff period}\label{watstiff}

After the end of fast-roll inflation the waterfall field is released and runs
down the potential slope, giving rise to a stiff period. We can approximate the
scalar potential in Eq.~(\ref{Vphi}) when \mbox{$\phi>M$} as
\begin{equation}
  V(\phi)\simeq\frac{\frac14\lambda M^4}{\left[\frac12\exp(\phi/M)\right]^4}
  =4\lambda M^4\exp(-4\phi/M)\,.
\label{Vkin}
\end{equation}  

A canonical scalar field rolling down an exponential potential of the form
\mbox{$V\propto\exp(-\kappa\phi/m_P)$} soon assumes
an attractor solution, which corresponds to equation of state (barotropic)
parameter given by \mbox{$w_\phi=-1+\kappa^2/3$} (provided \mbox{$w_\phi\leq 1$})
\cite{powerlaw,copexp,mybook}.
Thus, in our case, we expect the rolling waterfall field to be characterised by
the barotropic parameter
\begin{equation}
w_\phi=-1+\frac{16}{3}\left(\frac{m_P}{M}\right)^2\;\Leftrightarrow\;
M=\frac{4\,m_P}{\sqrt{3(1+w_\phi)}}\,.
\label{w}
\end{equation}

The stiff period is therefore not free-fall, with \mbox{$w_\phi=1$} and the
field is not fully dominated by its kinetic energy, because in the
exponential attractor evolution, all the terms of the Klein-Gordon equation of
motion are comparable. However, the barotropic parameter can still be larger
than 1/3.

A stiff period when \mbox{$1/3<w_\phi\leq 1$} results in a peak in the
spectrum of gravitational waves, which are produced by primordial inflation
\cite{sahni}. This peak corresponds to frequencies which re-enter the horizon
during the stiff period. Now, in kination when \mbox{$w_\phi=1$}, this spike is
very sharp and corresponds to high frequencies, beyond observational
capabilities in the foreseeable future. If kination lasted longer, so that lower
frequencies of gravitational waves can still re-enter the horizon during
kination, then the peak becomes too pronounced and affects the process of
Big Bang Nucleosynthesis (BBN) \cite{haroBBN}.

However, if $w_\phi$ is less than unity but still larger than the
radiation value of 1/3, then the peak corresponding to the stiff period is
milder. Then the stiff period can last longer, allowing primordial gravitational
waves of lower frequencies to be enhanced without threatening BBN. In
Ref.~\cite{FigTanin}
it was shown that in the range \mbox{$0.46\leq w_\phi\leq 0.56$},
detectable gravitational wave frequencies are amplified such that they will be
observable in the near future by LISA, without disturbing BBN. In view
of Eq.~(\ref{w}), this range corresponds to the range
\mbox{$1.85\leq M/m_P\leq 1.91$}, i.e. \mbox{$M\simeq 2m_P$}.
Using Eq.~(\ref{Malpha}), we find \mbox{$0.57\leq\alpha\leq 0.61$}.

In the following, to help with our analytic treatment, we choose
\mbox{$w_\phi=1/2$} (\mbox{$\alpha\approx 0.6$}),
which corresponds to \mbox{$M=\frac{4\sqrt 2}{3}\,m_P$}.
Using this value in Eq.~(\ref{lambda}), we find
\mbox{$\lambda\simeq3.2\times 10^{-11}$}. Then, Eq.~(\ref{NFR}) suggests that 
\mbox{$N_{\rm FR}=13.47$}. Note that standard Coleman-Weinberg hybrid inflation
in supergravity, when \mbox{$V(\sigma)\propto\ln\sigma$}, is brought into
agreement with Planck observations if there is a bout of inflation subsequent
to primordial inflation \cite{mine}.

\section{Reheating}

The density during the stiff matter period scales as
\mbox{$\rho_\phi\propto a^{-3(1+w_\phi)}=a^{-9/2}$}, where we
used the approximation \mbox{$w_\phi\approx 1/2$}. Thus, the density parameter
of radiation during the stiff period is
\begin{equation}
\Omega_r=\frac{\rho_r}{\rho_\phi}\propto\frac{a^{-4}}{a^{-9/2}}=\sqrt a\,.
\label{Omega}
\end{equation}
Therefore, we obtain
\begin{equation}
  1\sim\Omega_r^{\rm reh}=\Omega_r^{\rm end}\sqrt{\frac{a_{\rm reh}}{a_{\rm end}}}
  \;\Rightarrow\;
  \frac{a_{\rm end}}{a_{\rm reh}}\simeq\left(\Omega_r^{\rm end}\right)^2\,,
\label{aratio}
\end{equation}  
where `end' denotes the end of fast-roll inflation and `reh' denotes the moment
of reheating, when radiation becomes dominant and the Hot Big Bang begins.
Using the above, we find
\begin{equation}
\left.\begin{array}{l}
  \rho_r^{\rm end}\simeq\Omega_r^{\rm end}\rho_\phi^{\rm end}\\
  \rho_r^{\rm reh}=\rho_r^{\rm end}\left(\frac{a_{\rm end}}{a_{\rm reh}}\right)^4
\end{array}\right\}\Rightarrow\;
\rho_\phi^{\rm reh}\simeq\rho_\phi^{\rm end}\left(\Omega_r^{\rm end}\right)^9\,.
\end{equation}  
Under the simplifying assumption that
\mbox{$\rho_\phi^{\rm end}\simeq\rho_\phi^{P/T}\simeq V_{\rm inf}\sim 10^{-10}m_P^4$},
we can estimate the reheating temperature
\begin{equation}
  T_{\rm reh}\simeq\left(\frac{30}{\pi^2 g_*}\right)^{1/4}
  \left(\Omega_r^{\rm end}\right)^{9/4}\times 10^{-5/2}\,m_P\;,
\label{Treh}  
\end{equation}
where \mbox{$g_*\lsim {\cal O}(100)$} is the number of effective relativistic
degrees of freedom and we used that
\mbox{$\rho_r^{\rm reh}=\frac{\pi^2}{30}g_*T_{\rm reh}^4$}.

In Ref.~\cite{FigTanin} it is shown that we need
\mbox{1~MeV$\,\lsim T_{\rm reh}\lsim 150\,$MeV} for observable primordial
gravitational waves.\footnote{Note that the lower bound on the reheating
  temperature is about 4~MeV \cite{4MeV}.} Considering
\mbox{$T_{\rm reh}\sim 10^2\,$MeV}, Eq.~(\ref{Treh}) suggests that
\mbox{$\Omega_r^{\rm end}\sim 10^{-8}$}. It is easy to show that either
gravitational particle production, or even the outburst of tachyonic
fluctuations at the phase transition are not enough to generate the desired
reheating efficiency.\footnote{Recall that there is a period of
  \mbox{$N_{\rm FR}\simeq 13$} e-folds of fast-roll inflation after the phase
  transition, which dilutes significantly the products of the tachyonic particle
  production.}
Therefore, another mechanism is needed for reheating. As an example, we
employ Ricci reheating \cite{Riccimine,Riccireh,Riccilast},
which has the advantage of not introducing any
additional coupling of the infaton or the waterfall to the spectator field
responsible for reheating (in contrast to other mechanisms, such as instant
preheating \cite{instant}). It also does not depend on initial conditions
(as does curvaton reheating, for example \cite{curvreh,curvrehmine}).%
\footnote{For other reheating mechanisms in NO inflation see
  Refs.~\cite{haroreh,sasakireh,warmQI,dalianis}.} This is why it has been
considered when modelling quintessential inflation (e.g. see Ref.~\cite{ours}).

Ricci reheating considers a non-minimally coupled scalar field $\chi$, with
Lagrangian density
\begin{equation}
  {\cal L}_\chi=\frac12(\partial\chi)^2-\frac12\xi R\chi^2+\cdots\,,
\label{Lchi}
\end{equation}
where \mbox{$(\partial\chi)^2=-\partial_\mu\chi\,\partial^\mu\chi$},  $R$ is the
Ricci scalar, $\xi$ is the non-perturbative non-minimal coupling to gravity and
the ellipsis denotes higher order terms, which can stabilise the potential of
$\chi$. The Ricci scalar is \mbox{$R=3(1-3w)H^2$}, where $w$ is the barotropic
parameter of the Universe. During inflation, both primordial and fast-roll, we
have \mbox{$w=-1$}, which means that \mbox{$R=12H^2$} and the non-minimal
coupling generates a positive effective mass squared for the $\chi$ field.
After the end of fast-roll inflation we have the waterfall stiff period 
with \mbox{$w=w_\phi=1/2$}. As a result, \mbox{$R=-\frac32 H^2$}
and the effective mass squared of $\chi$ becomes tachyonic. Consequently, there
is a tachyonic outburst of $\chi$-particles, which eventually decay into the
radiation bath of the Hot Big Bang.

Let us estimate the reheating efficiency $\Omega_r^{\rm end}$ of the process.
The density of the produced radiation at the phase transition is roughly
\mbox{$\rho_r^{\rm end}\simeq\frac12|m_\chi^2|\langle\chi^2\rangle$}, where
\mbox{$m_\chi^2=-\frac32\xi H^2$} is the effective mass-squared of the
$\chi$-field and \mbox{$\langle\chi^2\rangle\simeq |m_\chi^2|$} is its
expectation value (squared) at the phase transition. Thus,
\mbox{$\rho_r^{\rm end}\simeq\frac12|m_\chi^2|^2\simeq\frac98\xi^2H_{\rm end}^4$}.
Then, for the reheating efficiency we find
\begin{equation}
  \Omega_r^{\rm end}\simeq\frac{\rho_r^{\rm end}}{\rho_\phi^{\rm end}}\simeq
  \frac{\frac98\xi^2H_{\rm end}^4}{3H_{\rm end}^2m_P^2}=\frac38\xi^2
  \left(\frac{H_{\rm end}}{m_P}\right)^2\,.
\end{equation}  
During fast-roll inflation, the Hubble parameter is roughly constant so that
\mbox{$H_{\rm end}^2\simeq H_{P/T}^2$}, which is given by Eq.~(\ref{HPT}). Using
the selected value of \mbox{$M=\frac{4\sqrt 2}{3}\,m_P$} (such that
\mbox{$w_\phi=1/2$}) we obtain
\mbox{$\Omega_r^{\rm end}\simeq\frac{32}{81}\,\xi^2\lambda$}. Demanding that
\mbox{$\Omega_r^{\rm end}\sim 10^{-8}$} and using Eq.~(\ref{lambda}) we find
that \mbox{$\xi\simeq 30$}.

\section{The peak in the spectrum of gravitational waves}

The background of primordial gravitational waves generated during inflation
(primordial and/or fast-roll) acquires a spectrum given by \cite{latekination}
\begin{equation}
\Omega_{\rm GW}(f)\propto f^\beta\quad{\rm where}\quad
  \beta=-2\left(\frac{1-3w}{1+3w}\right)\,,
  \label{beta}
\end{equation}  
where $f$ is the frequency and $w$ is the barotropic parameter of the Universe.
For the modes which re-enter the horizon during the stiff period
\mbox{$f_{\rm reh}<f<f_{\rm end}$} when \mbox{$w=w_\phi=1/2$} we have
\mbox{$\beta=2/5$}. Then, the gravitational wave spectrum is
\begin{equation}
  \Omega_{\rm GW}(f)\simeq \Omega_{\rm GW}^{\rm rad} 
  \times
  \left\{\begin{array}{ll}
  (f/f_{\rm reh})^{2/5} & f_{\rm reh}<f<f_{\rm end}\\
  1 & f_{\rm eq}<f<f_{\rm reh}\\
  (f_{\rm eq}/f)^2 & f_0<f<f_{\rm eq}
\end{array}\right.\;,
  \label{OmegaGW}
\end{equation}
where `eq' denotes the time of equal radiation and matter densities (equality)
and `0' denotes the present. In the above, $\Omega_{\rm GW}^{\rm rad}$ is
a constant which we evaluate below, where with `rad' we denote the
modes which re-enter the horizon during the radiation era.

The characteristic frequencies above can be estimated as follows. For a given
momentum scale $k$, the corresponding frequency is \cite{latekination}
\begin{equation}
f=\frac{H_k}{2\pi}\frac{a_k}{a_0}\,,
  \label{fk}
\end{equation}
where the subscript `$k$' denotes the time when the scale in question re-enters
the horizon after inflation.

In the case of $f_{\rm end}$ we find
\mbox{$f_{\rm end}=(H_{\rm end}/2\pi)(a_{\rm end}/a_0)$}. Now, we have
\begin{equation}
  \frac{a_{\rm end}}{a_0}\simeq\frac{T_0}{T_{\rm end}}\sim
  \frac{T_{_{\rm CMB}}}{(\rho_r^{\rm end})^{1/4}}\sim
  \frac{T_{_{\rm CMB}}}{10^{-2}\rho_{\rm end}^{1/4}}\sim 10^{-27}\,,
\end{equation}  
where we considered that \mbox{$T_{_{\rm CMB}}\sim 10^{-13}\,$GeV},
\mbox{$\rho_r^{\rm end}=\Omega_r^{\rm end}\rho_{\rm end}$} with
\mbox{$\Omega_r^{\rm end}\sim 10^{-8}$} and
\mbox{$\rho_{\rm end}\sim 10^{-10}\,m_P^4$}.
Using that \mbox{$H_{\rm end}\sim 10^{-5}\,m_P$}, we find
\begin{equation}
  f_{\rm end}\sim 10^{-14}\,{\rm GeV}\sim 10^{10}\,{\rm Hz}\,.
\label{fend}
\end{equation}

For $f_{\rm reh}$ we consider that (cf. Eq.~(\ref{fk}))
\begin{equation}
  \frac{f_{\rm end}}{f_{\rm reh}}=
  \frac{H_{\rm end}}{H_{\rm reh}}\frac{a_{\rm end}}{a_{\rm reh}}\sim
  \left(\frac{t_{\rm reh}}{t_{\rm end}}\right)^{5/9}\sim
  \left(\frac{a_{\rm reh}}{a_{\rm end}}\right)^{5/4}\sim(\Omega_r^{\rm end})^{-5/2}\sim
  10^{20}\,,
  \label{20}
\end{equation}
where we used Eq.~(\ref{aratio})  and that, during the stiff period we have
\mbox{$a\propto t^{2/3(1+w)}=t^{4/9}$}, with \mbox{$H\propto t^{-1}$},
Therefore, from Eqs.~(\ref{fend}) and (\ref{20}) we obtain
\begin{equation}
  f_{\rm reh}\sim 
  10^{-34}\,{\rm GeV}\sim 10^{-10}\,{\rm Hz}\,.
\label{freh}
\end{equation}

For $f_{\rm eq}$ we find (cf. Eq.~(\ref{fk}))
\begin{equation}
  f_{\rm eq}=\frac{H_{\rm eq}}{2\pi}\frac{a_{\rm eq}}{a_0}\simeq
  \frac{\sqrt{\rho_{\rm eq}}}{2\pi\sqrt 3\,m_P}
\left(\frac{t_{\rm eq}}{t_0}\right)^{2/3}\sim 10^{-5}\,\frac{T_{\rm eq}^2}{m_P}\,,  
\end{equation}
where we ignored dark energy and considered that \mbox{$t_{\rm eq}\sim 10^4\,$y}
and \mbox{$t_0\sim 10^{10}\,$y}. Using that
\mbox{$T_{\rm eq}\sim 1\,$eV}, we obtain
\begin{equation}
f_{\rm eq}\sim 10^{-41}\;{\rm GeV}\sim 10^{-17}\,{\rm Hz}\,.
\label{feq}
\end{equation}

Finally, for $f_0$ we readily find \mbox{$f_0=H_0/2\pi$} (cf. Eq.~(\ref{fk}))
so that
\begin{equation}
  f_0\sim 10^{-43}\,{\rm GeV}\sim 10^{-19}\,{\rm Hz}\,,
\label{f0}
\end{equation}  
where we used that \mbox{$H_0\sim 10^{-33}\,$eV}.

In order to estimate $\Omega_{\rm GW}^{\rm rad}$
in Eq.~(\ref{OmegaGW}) we need to calculate
the density parameter of gravitational radiation at present. We find
\begin{equation}
  \Omega_{\rm GW}^0=\frac{\rho_{\rm GW}^0}{\rho_0}\simeq
  \frac{\rho_{\rm GW}^{\rm end}\left(\frac{a_{\rm end}}{a_0}\right)^4}%
{\rho_{\rm end}\left(\frac{a_{\rm end}}{a_{\rm reh}}\right)^{9/2}
\left(\frac{a_{\rm reh}}{a_{\rm eq}}\right)^4
\left(\frac{a_{\rm eq}}{a_0}\right)^3}=
\Omega_{\rm GW}^{\rm end}\left(\frac{a_{\rm end}}{a_{\rm reh}}\right)^{-1/2}
\frac{a_{\rm eq}}{a_0}\,,
\end{equation}  
where, during the stiff period, \mbox{$\rho\propto a^{-3(1+w)}=a^{-9/2}$}.
Using the fact that, at the end of fast-roll inflation we have
\mbox{$\rho_{\rm GW}^{\rm end}\sim H_{\rm end}^4$} we find
\mbox{$\Omega_{\rm GW}^{\rm end}\sim
  \frac{H_{\rm end}^4}{H_{\rm end}^2m_P^2}\sim 10^{-10}$}, where
\mbox{$H_{\rm end}\sim 10^{-5}\,m_P$}. Then, in view of Eq.~(\ref{aratio}),
the above becomes
\begin{equation}
\Omega_{\rm GW}^0\sim 10^{-14}/\Omega_r^{\rm end}\sim 10^{-6}\,.
\label{OmegaGW0} 
\end{equation}

Now, in view of Eq.~(\ref{OmegaGW}), we have
\begin{eqnarray}
  &  &
  \Omega_{\rm GW}(f)\equiv\frac{{\rm d}\Omega_{\rm GW}}{{\rm d}\ln f}\nonumber\\
\Rightarrow & &
\Omega_{\rm GW}^0=\int_{f_0}^{f_{\rm end}}\Omega_{\rm GW}(f)\frac{{\rm d}f}{f}\simeq
\frac52\,\Omega_{\rm GW}^{\rm rad}
\left(\frac{f_{\rm end}}{f_{\rm reh}}\right)^{2/5}\sim\Omega_{\rm GW}^{\rm rad}
\times 10^8 \nonumber\\
\Rightarrow && \Omega_{\rm GW}^{\rm rad}\sim 10^{-14}\,,
\label{A}
\end{eqnarray}
where we considered Eqs.~(\ref{fend}), (\ref{freh}) and (\ref{OmegaGW0})
and also that the integral is dominated by the high-frequency part.
It is important to note here that the BBN bound is an integrated constraint.

\begin{figure} [t]
\vspace{-6cm}

\centering
\mbox{\hspace{-1cm}
\includegraphics[width=20cm]{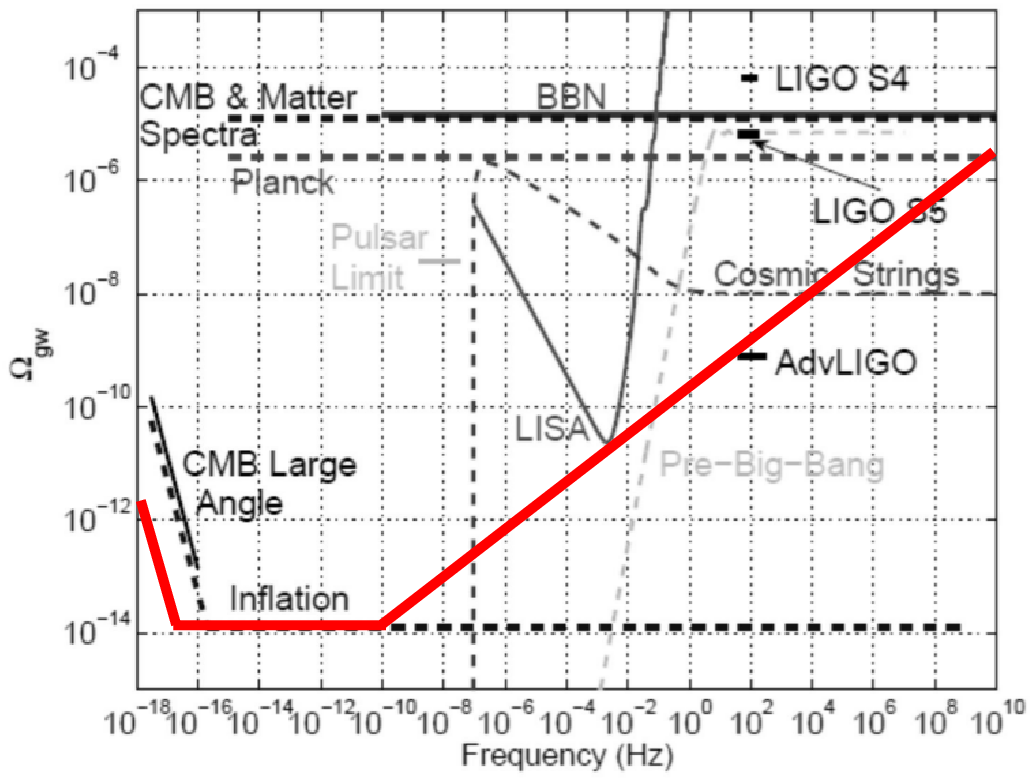}}
\vspace{-15.5cm}
\caption{The solid red line depicts the spectrum of primordial gravitational
  waves in our scenario, with stiff barotropic parameter
  \mbox{$w\approx\frac12$} and reheating temperature
  \mbox{$T_{\rm reh}\sim 10^2\,$MeV}.
  The peak in the spectrum of the gravitational waves is almost saturating the
  BBN bound, depicted by the horizontal solid black line.
  In the figure, the expected observational capability of Advanced LIGO and LISA
  are shown. It is evident that our scenario produces marginally observable
  (by LISA and Advanced LIGO) primordial gravitational waves. The observatinal
  bounds have been taken from Ref.~\cite{GWobs}.}  
\label{GWplot}
\end{figure}

\begin{figure} [t]
\vspace{-6cm}

\centering
\mbox{\hspace{-1cm}
\includegraphics[width=20cm]{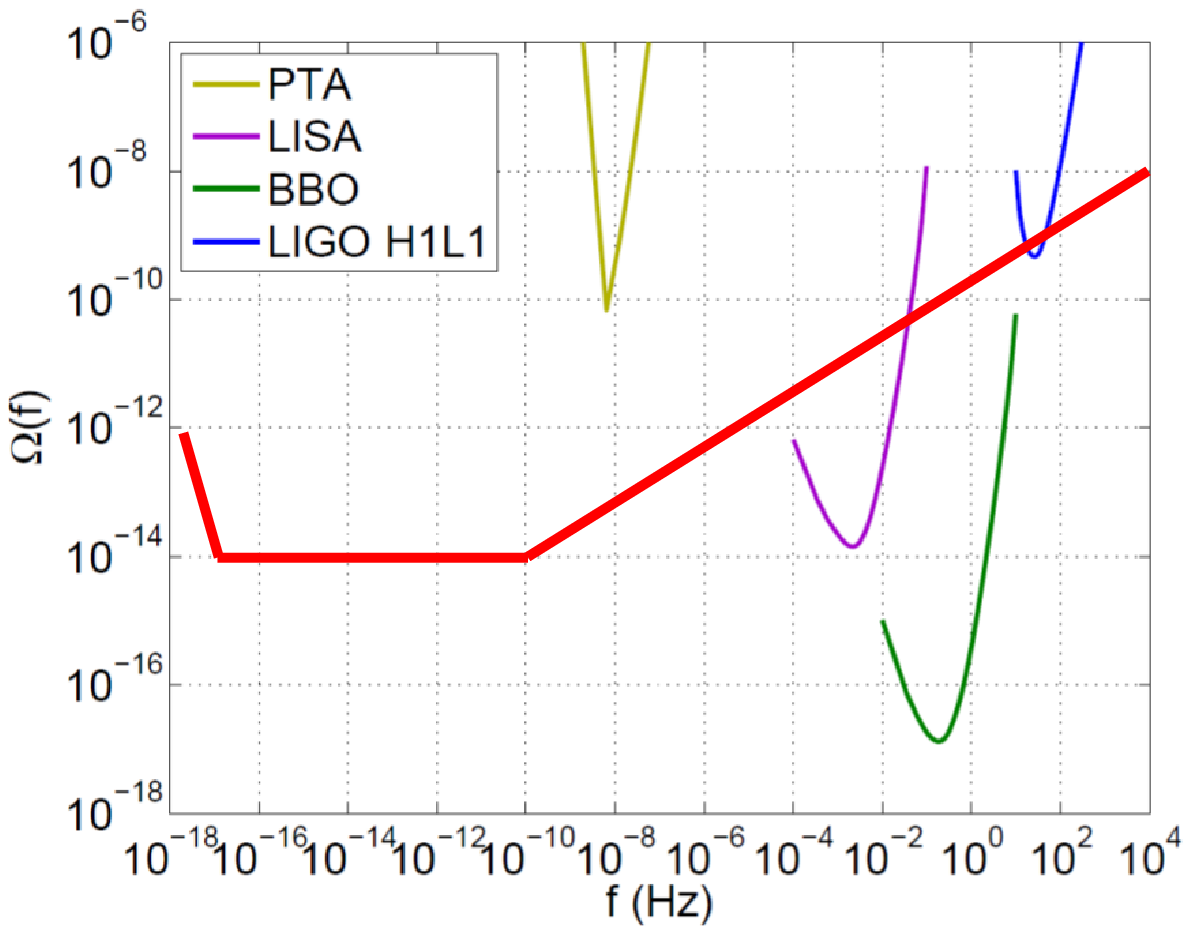}}
\vspace{-15.5cm}
\caption{The solid red line depicts the spectrum of primordial gravitational
waves in our scenario, with stiff barotropic parameter
\mbox{$w\approx\frac12$} and reheating temperature
\mbox{$T_{\rm reh}\sim 10^2\,$MeV}.
  The estimated observational capability of Advanced LIGO, LISA and BBO
  are shown. The predicted primordial gravitational radiation in our scenario
  is well observable by LISA and the Hanford-Livingston (H1L1) pair of
  Advanced LIGO detectors and even more so by BBO. The latter might also be able
  to discern the distinct slope in the gravitational wave spectrum from the
  flat spectrum expected by minimal inflation. 
  The bounds have been taken from Ref.~\cite{GWobs}.}  
\label{waterfig}
\end{figure}

We plot Eq.~(\ref{OmegaGW}) with \mbox{$\Omega_{\rm GW}^{\rm rad}\sim 10^{-14}$} in
Fig.~\ref{GWplot}, using also
Eqs.~(\ref{fend}), (\ref{freh}), (\ref{feq}) and (\ref{f0}). It is evident
that the peak in the gravitational wave spectrum is marginally observable
by~LISA and within reach of Advanced LIGO.%
\footnote{Here we are interested in order of magnitude estimates, but it
  must be noted that the spectral density of gravitational waves
  $\Omega_{\rm GW}(f)$ is also mildly sensitive to the effective degrees of
  freedom $g_*$ \cite{GWg*}.}
In other estimates, the sensitivity of LISA is larger after gathering enough
data. As a result, as shown in Fig.~\ref{waterfig}, the peak of primordial
gravitational waves is well within observability by LISA.
Also, as depicted in Fig.~\ref{waterfig}, the enhanced primordial gravitational 
waves could be clearly seen by other future missions, such as BBO \cite{BBO}.

We can estimate the density parameter of gravitational waves during
Big Bang Nucleosynthesis (BBN) as follows. Today, the density parameter of
radiation is \mbox{$\Omega_r^0\sim 10^{-4}$}. Thus we find
\begin{equation}
  \left.\frac{\rho_{\rm GW}}{\rho_r}\right|_0=
  \left.\frac{\Omega_{\rm GW}}{\Omega_r}\right|_0\sim 10^{-2}\,,
\end{equation}
where we used Eq.~(\ref{OmegaGW0}). Because
\mbox{$\rho_{\rm GW}, \rho_r\propto a^{-4}$} we have
\begin{equation}
  \Omega_{\rm GW}^{\rm BBN}\simeq\left.\frac{\rho_{\rm GW}}{\rho_r}\right|_{\rm BBN}
  =\left.\frac{\rho_{\rm GW}}{\rho_r}\right|_0\sim 10^{-2}\,,
\label{OmegaGWBBN}  
\end{equation}
where the sub/superscript `BBN' denotes the epoch of BBN. Thus, as expected
\mbox{$\Omega_{\rm GW}^{\rm BBN}\sim 10^{-2}$} saturates the bound from BBN, such
that the process is not disturbed by the primordial gravitational waves.


\section{Discussion}

We have investigated a generic model where there is an Enhanced Symmetry Point
(ESP) in field space where two flat directions are coupled. Of these, one flat
direction corresponds to the inflaton field, which drives primordial inflation
that resolves the fine-tuning problems of the Hot Big Bang and is responsible
for the generation of the curvature perturbation, which seeds the formation of
structure in the Universe. But we were interested in another aspect of
primordial inflation, namely the generation of gravitational waves. The other
flat direction can be a modulus field because it has a Planckian Vacuum
Expectation Value (VEV). The coupling between the two gives rise to the standard
hybrid mechanism \cite{hybrid}, which terminates primordial inflation via a
phase transition, when the inflaton field reaches near the ESP. After the phase
transition, the system rolls along the waterfall direction, sliding off from
the central potential hill, and driving a period of fast-roll hilltop inflation
\cite{fastroll}. So far, the scenario does not differ much from many such
configurations considered in the literature.

Things change when considering that the kinetic term of the waterfall field is
non-trivial and features a pole at the VEV, in the manner of
$\alpha$-attractors \cite{alpha}.
To assist our intuition, we switch to the canonically normalised waterfall
field, for which the VEV is displaced at infinity. The hybrid scenario is not
affected because near the ESP (at the top of the potential hill), the
non-canonical waterfall field is approximately canonical. However,
as the waterfall field slides away from the ESP, its potential is deformed and,
instead of rushing towards its VEV after the end of hilltop fast-roll inflation,
it follows an exponential attractor solution. This solution suggests that the
equation of state of the Universe is delicately dependent on the waterfall VEV
$M$. If \mbox{$M\simeq 2m_P$} then the resulting equation of state is such that
it drives a stiff period, with a barotropic parameter \mbox{$w\approx 1/2$}.
The significance of this, is that there is a peak in the spectrum of primordial
gravitational waves, such that they can be observable in the near future by LISA
\cite{FigTanin}.

The fact that a peak in the spectrum of gravitational waves is generated when
the Universe
after inflation (which produces them) enters a stiff period, is well known
\cite{sahni}.
However, most models which result in such a period consider a stiff phase with
\mbox{$w=1$}, dominated by the kinetic energy density of the inflaton field.
This is why this period is called kination \cite{kination}.
In our case, the exponential
attractor solution is such that the potential and kinetic energy densities of
the waterfall field are comparable, so this waterfall stiff period is not really
a period of kinetic energy density domination. The significance of this is as
follows.

In kination (with \mbox{$w=1$}), the peak in the spectrum of primordial
gravitational waves is very sharp and located at too high frequencies to be
observable. If kination lasted long enough to approach observable frequencies,
then the peak would become so large that the total energy density in
gravitational waves would disturb Big Bang Nucleosynthesis (BBN) \cite{haroBBN},
one of the pillars of the Hot Big Bang. However, when \mbox{$w\approx 1/2$} the
peak inthe spectrum of gravitational waves is milder and so it can spread out to
frequencies
low enough to be observable without affecting BBN. We have demonstrated this in
our Fig.~\ref{GWplot}.

Thus, we find that, our setup can quite naturally generate
observable primordial gravitational waves. One only needs two flat directions
meeting at an ESP, with one of these having a non-canonical kinetic term with a
pole at its Planckian VEV $M$. The only tuning we require so-far is that
\mbox{$M\approx 2 m_P$}.

This is of course not enough. Achieving the largest possible growth in the
spectrum of the gravitational waves, which
leaves BBN unaffected, requires a reheating temperature about
\mbox{$T_{\rm reh}\sim 10^2\,$MeV} \cite{FigTanin}.
The outburst of tachyonic perturbations at
the phase transition (with original density $\sim H^4$), which terminates
primordial inflation and sends the waterfall field down its potential cannot
produce enough radiation to reheat the Universe this early. Thus, we have to
consider alternative reheating mechanisms. Many such mechanisms have been
considered in models of quintessential inflation, which feature a kination
period \cite{haroreh}. As an example, we employed the Ricci reheating
mechanism \cite{Riccimine,Riccireh,Riccilast},
which considers the influence of a non-minimal spectator scalar field, whose
effective mass squared changes sign at the end of hilltop fast-roll inflation
(not at the phase transition) producing an outburst of tachyonic perturbations
with original density $\sim\xi^2H^4$. We have shown that we obtain enough
radiation to achieve the desired reheating temperature when the non-minimal
coupling is \mbox{$\xi\simeq 30$}; a very reasonable value.

One possible criticism of the above scenario is that the excursion of the
canonical field is super-Planckian and this would result in radiative
corrections which could lift the flatness of the waterfall potential after the
end of the fast-roll hilltop inflation phase. A super-Planckian excursion of
the field might result also in a sizeable 5th-force problem, which could violate
the Principle of Equivalence. Of course, the interaction terms in the Lagrangian
density of the theory are of the form \mbox{$e^{\beta_i\varphi/m_P}{\cal L}_i$}
\cite{caroll},
where ${\cal L}_i$ is any gauge-invariant dimension-four operator (for example
for electromagnetism \mbox{${\cal L}_{\rm em}=-\frac14 F_{\mu\nu}F^{\mu\nu}$}), and
$\beta_i$ are some constants of order unity. Crucially, this expression features
the non-canonical waterfall field $\varphi$, whose excursion is only Planckian,
since its VEV is \mbox{$M\simeq 2m_P$}. Thus, we only need \mbox{$\beta_i\ll 1$}
to suppress radiative corrections and the 5th force problem. Still, one could
consider this as substantial fine tuning because the $\beta_i$ are many.
A better argument can be made if we take seriously the Ricci reheating
mechanism. In this mechanism, the thermal bath of the Hot Big Bang is solely due
to the decay products of the spectator field $\chi$, which is not coupled to
either the inflaton $\sigma$ or the waterfall field $\varphi$.\footnote{In fact,
$\chi$ could conceivably be the Higgs field itself.} As a result, we
may consider that both $\sigma$ and $\varphi$ are completely uncoupled to the
standard model and belong to a dark sector. As such, they do not cause any
violation to the Equivalence Principle.\footnote{A similar argument can be made
  for the curvaton reheating mechanism.} As far as the radiative corrections are
concerned, they could simply be responsible for generating the Planckian VEV
of $\varphi$.

Even so, it may be argued that, when the waterfall field approaches its
Planckian VEV, the perturbative form of the potential in Eq.~(\ref{Vvarphis})
is questionable. Firstly, the waterfall field $\varphi$ approaches is VEV
asymptotically, when for the canonical waterfall field
\mbox{$\phi\rightarrow\infty$}. This implies that any deformations of
the potential until reheating (afterwards it is negligible) are expected
to be mild. As such, the estimated values of the model parameters $M$, $\lambda$
and $\xi$ might be somewhat affected, but we do not expect this to be
substantial. 

Another potential issue has to do with the phase transition, which terminates
inflation and sends the waterfall field down the hilltop of its potential.
It can be argued that modelling the system as a rolling ball is not applicable
in this case, because the phase transition is non-perturbative. As such, the
validity of Eq.~(\ref{NFR}) could be undermined. However, investigating
the phase transition at the onset of fast-roll inflation in the appropriate
detail, produces very similar results as with our simple treatment here
\cite{fastrollmine}.\footnote{A recent study of the backreaction of waterfall
  fluctuations on the inflaton field at the phase transition, showed that the
  curvature of the potential in the inflaton direction must be substantial,
  while the strength of the ESP not too large (so $g\ll 1$), for the classical
  approximation to be valid \cite{mylova}. This depends on the choice of $g$
  and the inflaton potential $V(\sigma)$, which we assume to be such that
  the aforementioned backreaction is not strong.}
A related issue is of possible topological defects generated at the phase
transition. Firstly, there are \mbox{$N_{\rm FR}\simeq 13$} e-folds of fast roll
inflation following the phase transition, which would dilute somewhat any
topological defects. The kind of topological defects created has to do with the
nature of the waterfall field. For example, if $\varphi$ is complex, we expect
the formation of cosmic strings, which could be made harmless if they are
unstable.


\vspace{2cm}

\begin{figure} [h]
\vspace{-4.8cm}

\centering
\includegraphics[width=10.2cm]{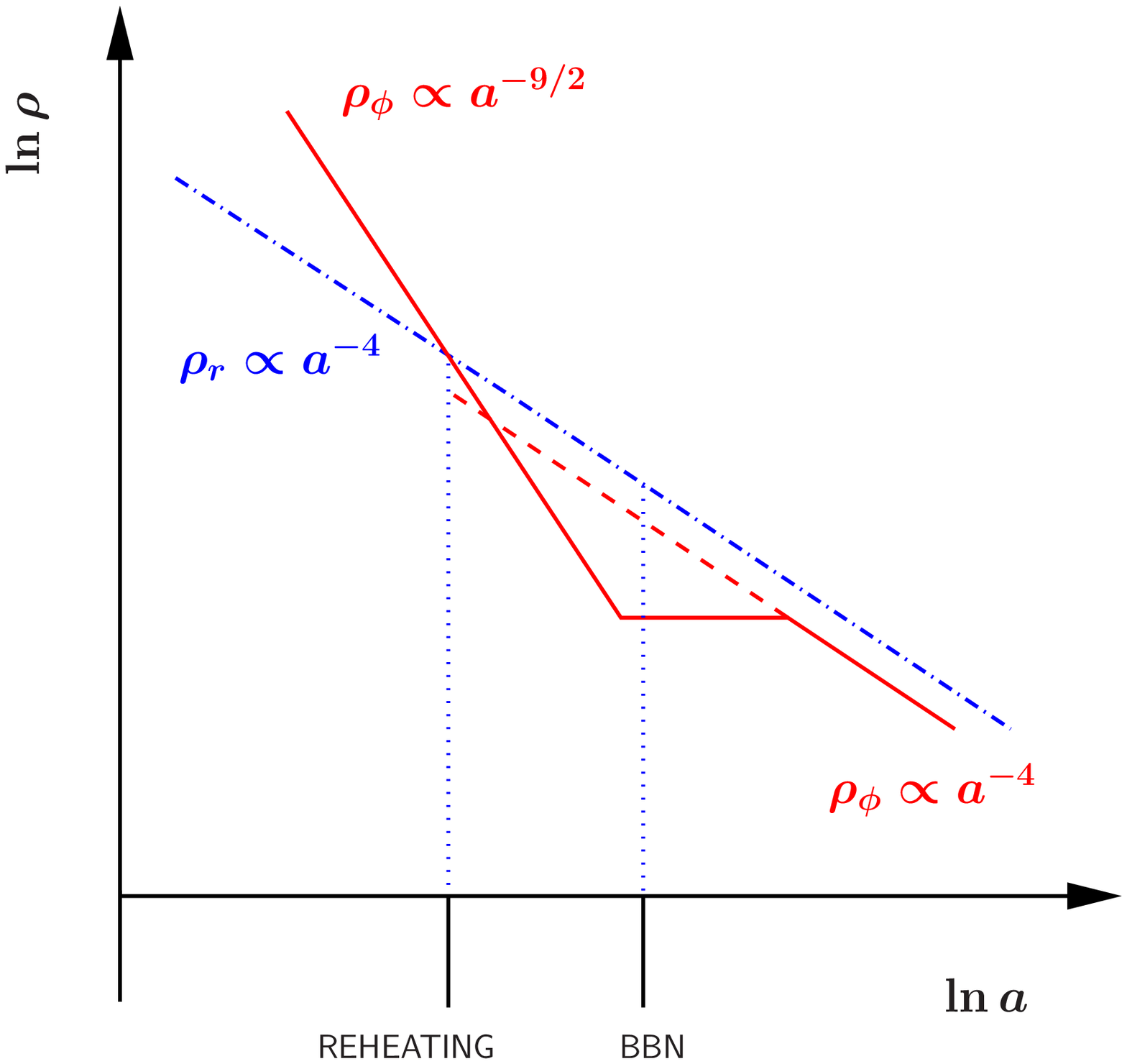} 
\vspace{-3.9cm}
\caption{Pictorial representation of how we expect that overshooting can
  temporarily decrease the contribution of the waterfall field to the density
  budget of the Universe, so that BBN remains undisturbed. In this log-log plot,
  the solid (red) line depicts the density of the scalar field while the
  dashed dot line (blue) depicts the density of the radiation thermal bath
  of the Hot Big Bang. Initially, the scalar field dominates the Universe
  and its density decreases as \mbox{$\rho_\phi\propto a^{-9/2}$}. The density of
  radiation decreases as \mbox{$\rho_r\propto a^{-4}$}. As a result, even though
  it is initially subdominant, radiation comes to dominate at the moment of
  reheating. When the scalar field becomes subdominant there is an attractor
  solution to its evolution which dictates that its density mimics the
  background at constant ratio. This attractor is depicted with dashed line
  (red). However, after reheating, the subdominant attractor is not immediately
  assumed. The scalar field is expected to overshoot the attractor, then become
  temporarily frozen until it can assume the attractor and continue rolling with
  density \mbox{$\rho_\phi\propto a^{-4}$}, mimicking the radiation background.
  Consequently, there is a brief period when the scalar field density is much
  smaller than the one which corresponds to the subdominant attractor evolution
  (note that in a log-log plot, substantial differences correspond to orders of
  magnitude). Because reheating occurs near BBN, 
  this temporary suppression of the contribution of the scalar field to the
  density budget of the Universe allows BBN not to be disturbed.
}  
\label{overshoot}
\end{figure}

Finally, let us consider what happens to the waterfall field after reheating.
Once the Universe becomes dominated by some substance other than the scalar
field, the exponential attractor changes and becomes such that the density of
the rolling scalar field mimics the background (whatever this is) and stays at
a constant ratio \cite{copexp,mybook}. In our case, this ratio is given
by \mbox{$\Omega_\phi=3/\kappa^2=\frac{3}{16}(\frac{M}{m_P})^2\approx\frac23$},
where \mbox{$M\approx\frac{4\sqrt 2}{3}\,m_P$}. Does this mean that we
cannot help but affect BBN after all, since the Universe content would contain
in effect an extra relativistic species? (the barotropic parameter of the scalar
field mimics that of the background, i.e. \mbox{$w_\phi=1/3$} during the
radiation era.) There is hope that we escape this danger, but only because
reheating occurs close to BBN.
Then, as the Universe expansion changes rate, we expect that the
scalar field would overshoot the subdominant attractor \cite{mybook}
and there will be some
limited period of time, when its contribution to the density budget of the
Universe is small enough to avoid disturbing BBN. This is possible only because
reheating is so close to BBN. Pictorially, this overshooting is depicted in
Fig.~\ref{overshoot}. Soon after BBN, the field is expected to assume the
subdominant exponential attractor. Therefore, in the matter era, we expect that
it comprises a large fraction of dark matter.
Needless to say that all the above warrant a detailed numerical investigation,
which we will do in a subsequent paper.

\section{Conclusions}

All in all, we have presented a toy-model of hybrid inflation, where the
waterfall field has a non-canonical kinetic term which features a pole at its
Planckian VEV. In this case, we have argued that, after inflation there is a
stiff period such that the corresponding primordial gravitational waves can be
enhanced enough to be observable by LISA and Advanced LIGO without disturbing
BBN.

\section*{Acknowledgements}
I am thankful to Samuel S\'anchez L\'opez for his help regarding
Fig.~\ref{3dplot} and to the anonymous referee, who made me realise that
the observability of the gravitational waves in the scenario presented is
greater than I originally thought.
This work was supported, in part, by the Lancaster-Manchester-Sheffield
Consortium for Fundamental Physics under STFC grant: ST/T001038/1.

\end{document}